# Chapter 4

# Semantic data discovery from Social Big Data

Bilal Abu-Salih[1], Pornpit Wongthongtham[2]
Dengya Zhu[3] , Kit Yan Chan[3] , Amit Rudra[3]

[1]The University of Jordan

[2] The University of Western Australia

[3] Curtin University

**Abstract:** Due to the large volume of data and information generated by a multitude of social data sources, it is a huge challenge to manage and extract useful knowledge, especially given the different forms of data, streaming data and uncertainty and ambiguity of data. Hence, there are still challenges in this area of BD analytics research to capture, store, process, visualise, query, and manipulate datasets to derive meaningful information that is specific to an application's domain. This chapter attempts to address this problem by studying Semantic Analytics and domain knowledge modelling, and to what extent these technologies can be utilised toward better understanding to the social textual contents. In particular, the chapter gives an overview of semantic analysis and domain ontology followed by shedding light on domain knowledge modelling, inference, semantic storage, and publicly available semantic tools and APIs. Also, the theoretical notion of Knowledge Graphs is reported and their interlinking with SBD is discussed. The utility of the semantic analytics is demonstrated and evaluated through a case study on social data in the context of politics domain.

## 4.1 Introduction

The evolution of Web technologies began with Web 0.0 in which the Internet was developed. Then it came to Web 1.0 which is considered as the static Web or the read-only Web where the information being presented in a form of the traditional document-centric way [1]. Then it evolved to Web 2.0 which is considered as the read-write Web where there are interaction and contribution between web users in a form of the user-generated contents or community-oriented information gathering. This era empowered the users with social media and blogs. The huge amount of data (known as big data) and unstructured content being generated on the Web makes a greatest challenge to users to process. Hence Web 3.0, known as the semantic executing Web, was developed to make the web content is understandable and process-able by computer. The underlying concept of the semantic web or the 'read-write-executing' web is that data are given semantic mark-up to define its meaning [2]. The semantic mark-up allows human web users to communicate with computerised applications. One of the challenges of presenting information on the web was that the web applications were not able to provide context to data and hence did not understand what was and what was not relevant [3]. The notion of semantic web enables software agents to understand and execute the web contents. Then it comes to Web 4.0, known as the mobile web, which is an alternate version where it connects all devices in the real and virtual world in real-time. The next web, known as the emotional web, which is still developing and its mode is still forming. The interaction between human web users and computers will be based on neurotechnology.

Social media plays a major role in our societies, reshaping the media arena, changing the rules of the game to a large extent, and providing ample space for expression and mutual dialogue. These forces have helped create platforms for the formation of public space or the public domain [4-7]. The new media have also weakened the role of the so-called gatekeeper and the role of the major traditional media (newspapers, radio, television, news outlets) in prioritising public opinion [8] [9-13]. It has the active ability to set and shape the nature of discourse and "Manufacturing Consent [14]". On the other hand, social media have been used excessively and irresponsibly by a certain slice of the community; their use has resulted in a state of political polarisation, ideological alignment, spreading rumours, and promoting extremism, racism and terrorism. The literature review emphasises the pivotal significance of understanding the contextual content of social data. This leads to a close acquaintance with users' beliefs and their attitudes [15-20]. Hence, improving the current technology tools and approaches in order to better understand the user's social content is inevitable. This leaves a room for a key challenge: how can we discover the domains of short text messages such as tweets? In particular, it is important to take into account the semantic relationships of terms in the user's textual content, particularly in short text messages such as tweets. For example, in the political domain, the term 'Labor', that is extracted from tweets, would be represented under the concept 'Political Party' in the "Politics" domain but would be a different concept in another domain such as the "Work" domain. This is a significant problem due to the brevity of tweets which prevents the machine from obtaining an accurate understanding of their textual content.

This chapter addresses this research problem and underlying issues. This chapter is organized as follow: Section 2 sheds light on Ontology and semantic analysis and the underlying technologies. Also, it gives a discussion on domain knowledge modelling, inference, storage and the related aspects. In sections 3 a set of semantic analysis tools and APIs are briefly presented. Section 4 introduces Knowledge Graphs and a discussion on selected applications that benefit from the integration of Knowledge Graphs and SBD is also provided. Section 7 reports on a case study incorporating semantic analysis to obtain better understanding to the social content in the context of politics domain. In the conclusion of this chapter a summary of the key aspects given in this chapter is provided, also the conclusion makes several recommendations for future research directions.

## 4.2 Semantic Analysis

### 4.2.1 Ontology: origin and definition

The term 'ontology' appears in two disciplines i.e. in a discipline of metaphysics and philosophical science and in a discipline of information science and artificial intelligence. Ontology, which is first introduced in a philosophical discipline, refers to the study of being or existence as well as the basic categories [21]. Ontology explains the nature and essential properties and relations between all beings. Modelling ontology is considered as a meta-model, which is a model of a model, describing the model on a higher layer. In the latter part of the twentieth century, researchers in the artificial intelligence discipline have become active in computational modeling of ontologies that would deliver automated reasoning capabilities. Tom Gruber generated expansive interest across the computer science community by defining ontology as "an explicit specification of a conceptualisation" [22]. Conceptualization is the formulating of domain knowledge. It is an abstract model of a phenomenon. The specification is the representation of the conceptualizations in a concrete form [23]. The specification will lead to commitment in semantic structure. In short, an ontology is the working model of entities. Notably there is development of new software tools to facilitate ontology engineering. Ontology engineering is an effort to formulate an exhaustive and rigorous conceptual schema within a given domain. Basically, ontology captures the domain knowledge through the defined concrete concepts (representing a set of entities), constraints, and the relationship between concepts, in order to provide a formal representation in machine understandable semantics. The purpose of ontology is to represent, share, and reuse existing domain knowledge.

### 4.2.2 Social media services incorporating semantic analysis

The use of ontology in the social media has been applied widely to infer semantic data in a broad range of applications. [24] presented an ontology-based, multi-agent solution for the wild animal traffic problem in Brazil. [25] and [26] both applied ontology to build applications in crisis situations. The former designed ontology for earthquake evacuation to help people find evacuation centers in earthquakes crises based on data posted in Twitter. The latter showed a geo-tagger that aims to process unstructured content and infer locations with the help of existing ontologies. [27] conducted a survey that addressed research issues related to processing socialmedia streams using semantic analysis. Some of the key questions which were the focus of this paper included: (i) How could Ontologies be utilized with Web of Data for semantically annotating social media contents? (ii) How could the annotation process discover hidden semantics in social media? (iii) How could trustworthiness of data be extracted from massive and noisy data? (iv) What are the techniques to model user identity in the digital world? (v) How could information retrieval techniques incorporate semantic analysis to retrieve highly relevant information? [28] presented an approach that helps mobile users in their search in the social networks by building an ontology-based context-aware module for mobile social networks. Their approach includes: 1. knowledge extraction from SN (implicit, explicit, (none) contextual data using API; 2. data cleansing; 3. knowledge modeling (knowledge of user's details and contextual information); 4. comparing user profiles and the contextual information; and 5. presenting retrieved data in mobile format. [29] proposed an approach intended to explore events from a twitter platform and to enrich an ontology designed for that purpose.

Statistical techniques have been used as another means of topic modeling and discovery in twitter mining. The two dominant statistical techniques that have been used are LDA (latent Dirichlet allocation) [30], and Latent Semantic Analysis (LSA). In LSA, an early topic modeling method has been extended to pLSA [31], which generates the semantic relationships based on a word-document co-occurrence matrix. LDA is based on an unsupervised learning model in order to identify topics from the distribution of words. These approaches have been widely used in sentiment analysis (Saif, He, and Alani 2011) and several modeling applications [32-37]. However, the high-level topics classifications that use these bag-of-words statistical techniques are inadequate and inferior [38]. The brevity and ambiguity of such short texts makes it more difficult to process topic modelling using these statistical models [39].

## 4.3 Domain knowledge modelling, inference and storage

### 4.3.1 Annotation and enrichment

The textual data can be semantically annotated with the concepts being defined in the domain ontologies. The annotation can then be enriched with a description of the concepts referring to the domain ontologies using knowledge graph and/or controlled vocabularies e.g. Dublin Core (DC4), Simple Knowledge Organization System (SKOS5), Semantically Interlinked Online Communities (SIOC6), etc. This allows each entity in the textual data to be specified with its semantic concept. The particular concepts can be further expanded into other related concepts and other entities instantiated by the concepts. The consolidation of this semantic information provides a detailed view of the entity captured in domain ontologies. The domain ontologies can be manipulated using Apache Jena API. Jena, which is a Java framework for building semantic web applications providing functionalities to create, read, and modify triples (subject–predicate–object) in ontologies.

### 4.3.2 Interlinking

Entities are interlinked with similar entities defined in other datasets to provide an extended view of the entities represented by the concepts. Equivalence interlinking specifies URIs (universal resource identifiers) that refer to the same resource or entity. Ontolgy Web Language provides support for equivalence links between ontology components and data. The resources and entities are linked through the 'owl#sameAs' relation; this implies that the subject URI and object URI resources are the same. Hence, the data can be explored in further detail. In the interlinking process, multiple vocabularies (i.e. Upper Mapping and Binding

Exchange Layer (UMBEL[1]), Freebase[2]– a community-curated database of well-known people, places, and things, YAGO[3] – a high-quality knowledge base, Friend-of-a-Friend10, DC[4], SKOS[5], SIOC[6], and DBPedia[7] knowledge base) are used to link and enrich the semantic description of resources annotated,

### 4.3.3 Semantic repository

A semantic repository represents a knowledge base which continues and updates the semantically rich annotated structured data. Ontology formalizes the conceptualized knowledge in a particular domain and provides explicit semantics by splitting concepts, their attributes, and their relationships from the instances. In the repository, there are terminological data that define concepts (classes), attributes (data properties), relationships (object properties), and axioms (constraints) as well as data that enumerates the instances (individuals). This enables different services support, such as concept based search, entailment to retrieve implied knowledge, instance-related information retrieval, etc. By using the semantic repository, query expansion for entity disambiguation can be performed to retrieve semantic description of entities.

In the repository, the structured data are stored as the RDF graph for persistence. Virtuoso (open source edition) triple store is used to store the RDF triples, ontologies, schemas, and expose it using a SPARQL endpoint. The SPARQL endpoint enables applications, users, software agents, and the like to access the knowledge base by posing SPARQL queries.

### 4.3.4 Politics ontology

Later in this chapter a case study on social political data analysis will be presented to incorporate IBM Watson Natural Language Understanding (NLU) and Politics ontology. The BBC produces a plethora of rich and diverse content about things that matter to the BBC's audiences, ranging from athletes, politicians, or artists [40]. The BBC uses domain Ontologies to describe the world and content the BBC creates and to manage and share data within the Linked Data platform. Linked Data provides an opportunity to connect the content together through various topics. Among the nine domain ontologies that the BBC has developed and uses, the Politics Ontology describes a model for politics, specifically in terms of local government and elections [41]. This was originally designed to cope with UK (England and Northern Ireland) Local, and European Elections in May 2014. The focus of the project is on Australian Politics; hence, we have developed a domain-specific Politics Ontology for Australian Politics by extending the BBC Politics

---

[1] https://lod-cloud.net/dataset/umbel
[2] https://developers.google.com/freebase
[3] http://www.foaf-project.org/
[4] https://www.dublincore.org/
[5] http://www.w3.org/2004/02/skos/
[6] http://sioc-project.org/
[7] https://wiki.dbpedia.org/

Ontology. We specified the ontology in Australian Politics having Australian politicians and Australian political parties. The concepts, instances, and relations are used in the annotation process. At this stage, the concept Politician has 53 instances of Australian politicians and the concept Political Party has 4 instances of Australian political parties. The politics ontology is being incrementally extended over time. Figure 1 shows the BBC Politics Ontology; Figure 2 shows the extended version of the BBC Politics Ontology using OntoGraf for visualization of the relationships in ontologies.

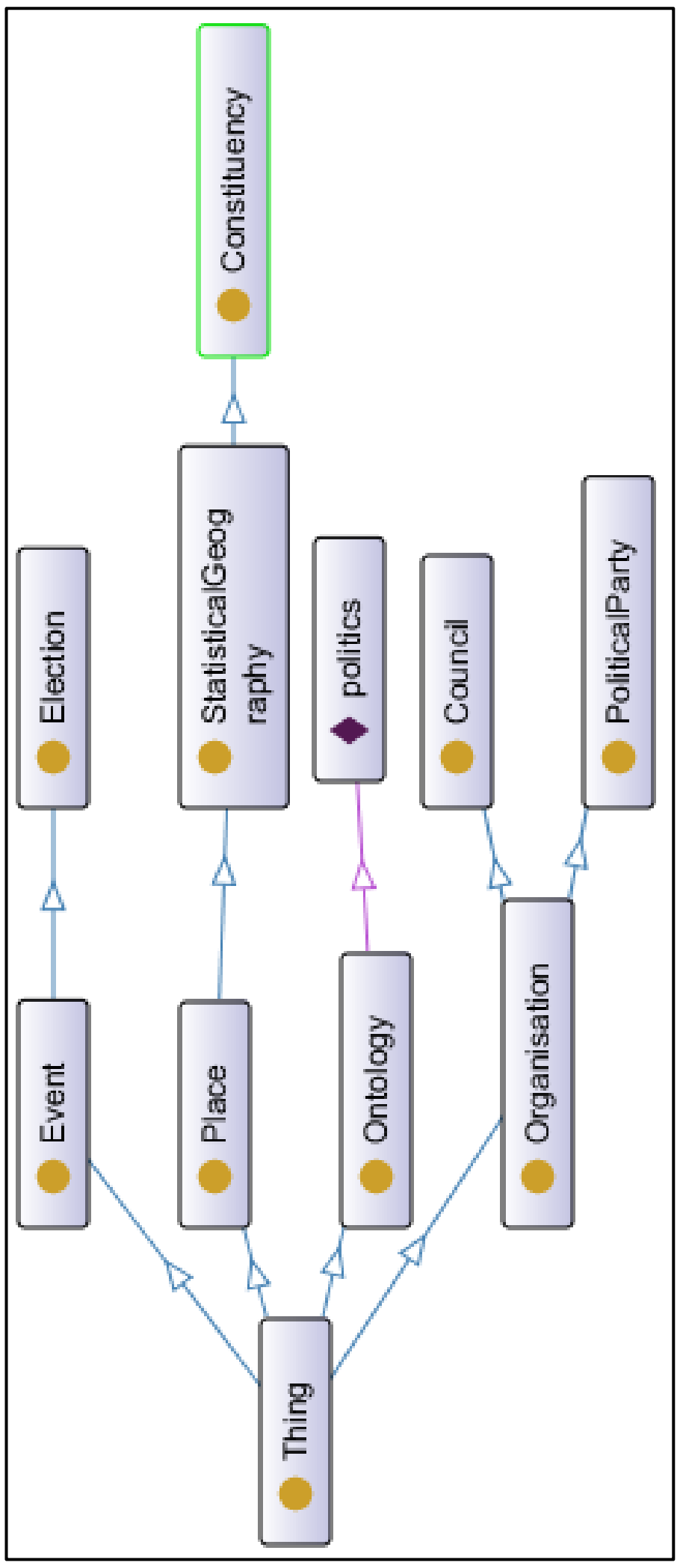

*Figure 1: BBC Politics ontology*

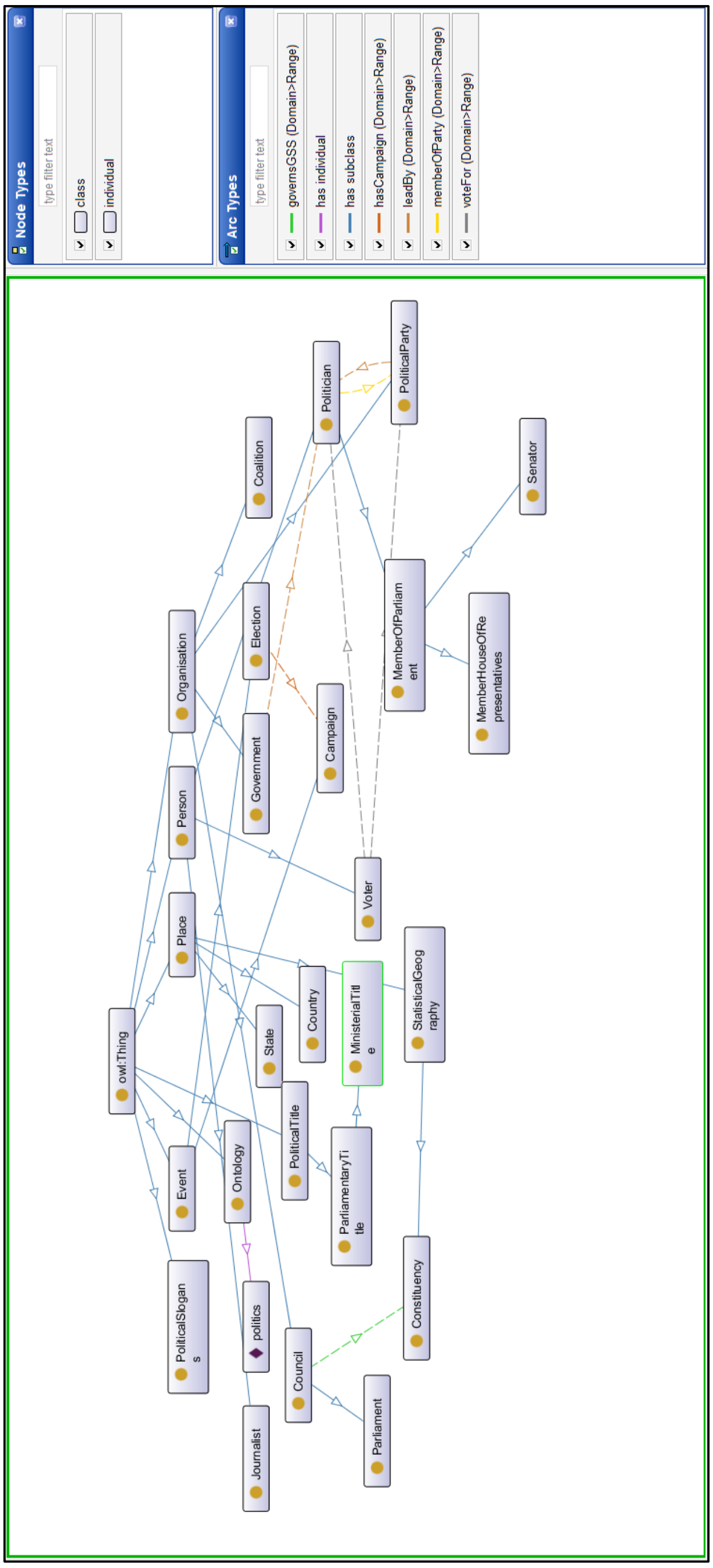

*Figure 2: BBC politics ontology extension*

In order to ensure the extended version of Politics Ontology is consistent, which is important as part of an ontology's development and testing, the Ontology needs to undergo a reasoning process. No reliable conclusion can be deduced otherwise. The extended version of the Politics Ontology has been reasoned to check its logical consistency using FaCT++, HermiT, Pellet, Pellet (Incremental), RacerPro, and TrOWL reasoners. The reasoners checked the class, object/data property hierarchies, the class/object property assertions, and whether there were the same individuals contained within the ontology. Consistency verification through a reasoner includes consistency checking, concept satisfiability, classification, and realization which are all standard inference services conventionally provided by a reasoner. The extended version of the Politics Ontology does not contain any contradictory facts.

## 4.4 Semantic Analysis tools and APIs for SBD

The results of the MIT survey [42] demonstrated the importance of data analytics as a means of elevating a company's position. Respondents indicated that they use data analytics to drive future strategies and day-to-day operations. However, some companies may be under the misconception that because there is data, this gives them fruitful results. In fact, the secret lies not in data collection, as decision makers may be swamped with more data, but in the acquisition of data that is relevant and meaningful [43]. In particular, a company must first decide which information in the collected data is essential in terms of, for example, its strategic objectives. Then, data analysis makes the difference and achieves the anticipated outcomes. These endeavours are fortified by collecting and utilizing the massive amount of data generated by OSNs. This, therefore, entails researchers and companies to build tools and APIs to carry out semantic analytics on textual content of social data in order to gain valuable insights. In this section, a brief on various open-source and commercial semantic analysis platforms and services are discussed.

### 4.4.1 IBM® Watson™ cognitive services

IBM Watson represents an ecosystem of a cohesive set of cognitive services that provide a variety of capabilities. Built on a top of NLP, Semantic Analysis, hypothesis generation and evaluation and dynamic learning, IBM Watson offers powerful multi-lingual solutions to address problems span form generic unstructured data analytics to industry-specific content [44, 45]. Amongst all products and services, IBM Watson Natural Language Understanding, formally known as Alchemy API, concerns with analyzing textual content and extract semantic features incorporating Semantic Web technologies that benefit SBD analytics. These semantic features include concepts, entities, emotion, keywords, relations, categories/taxonomies, semantic roles, sentiment analysis, etc. IBM Watson Understanding provides both domain-independent and customized domain analysis that can be imported implemented using their Knowledge Studio. Figure 3 illustrates the pipeline to carry out unstructured content using Watson Natural Language Understanding.

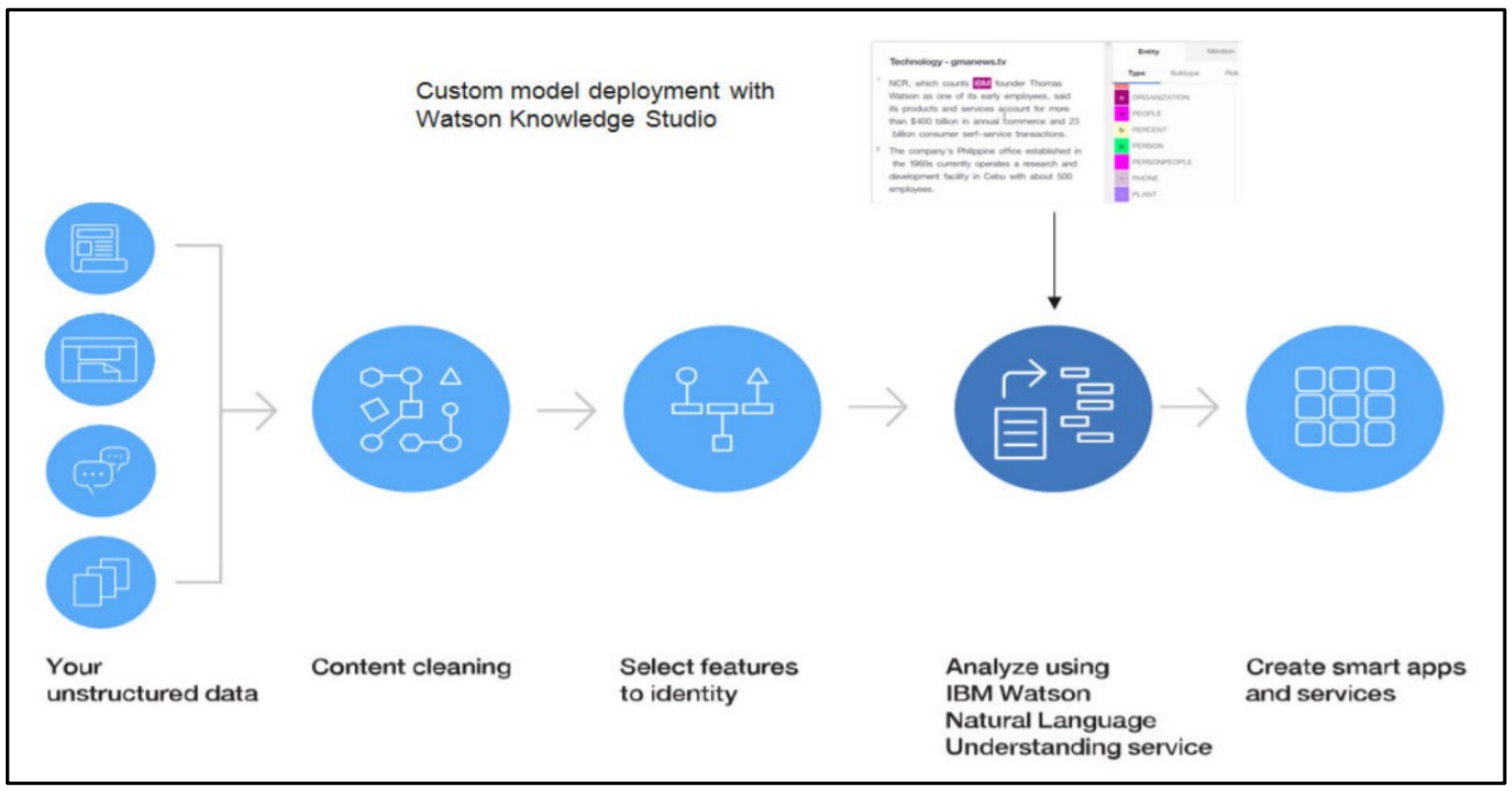

*Figure 3: IBM Watson Natural Language Understanding pipeline to analyse unstructured data [44]*

### 4.4.2 Lexalytics Intelligence Platform

Lexalytics uses cloud-based infrastructure to provide several services over textual contents [46]. Lexatics incorporate NLP, semantic analysis, machine learning and other technologies, and offer features such as, sentiment analysis, categorization, entity recognition, summarization and theme analysis. Lexalytics platform is a multi-lingual and can be accessed using both on-premise and RESTful API. Figure 4 illustrates Lexalytics intelligence platform. As depicted in the figure, Lexalytics embodies three key embedded engines, namely (i) Salience [47]: is a set of textual analytics and NLP libraries to be imported and implemented on-premise; (ii) Semantria API [48]: cloud-based solution for conducting NLP and textual analytics over RESTful API; and (iii) Semantria Storage and Visualization (SSV): a platform to assist in construction of intuitive dashboards to visualize patterns and trends over unstructured dataset. These three intrinsic platforms provide the capacity to monitor social media data, and various other un/semi-structured contents and have proven ability to gain actionable insights out of them [49, 50]. For example, listening to the Voice of Customer (VoC) and Voice of Employee (VoE) are amongst the core services provided by Lexalytics. The platform is able to gather, analyze and interpret unstructured social textual contents collected from, for example, customers and employees conversations on products, brands, people and services [51].

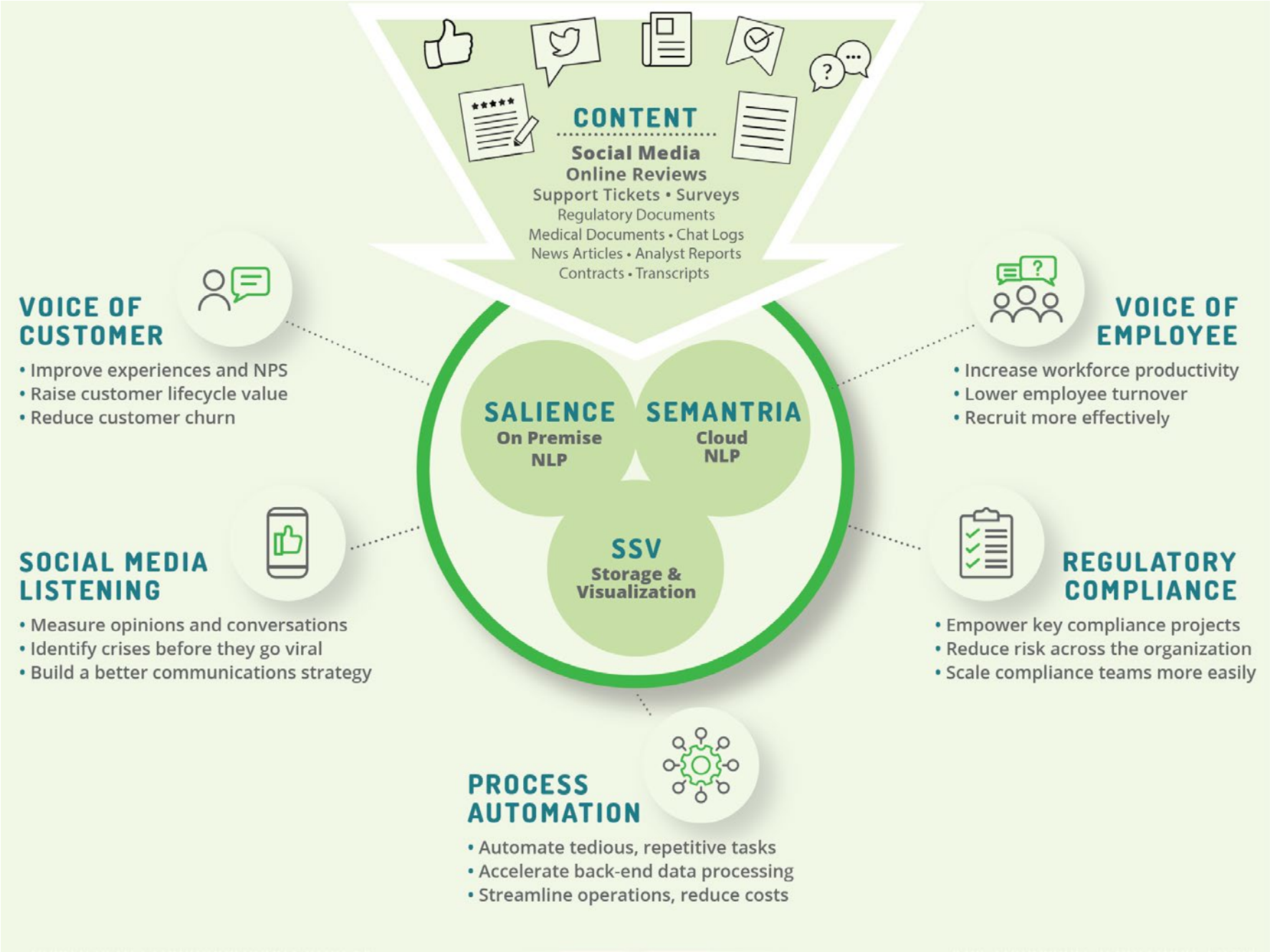

*Figure 4: Lexalytics Intelligence Platform[52]*

### 4.4.3 Cogito Discover ™

Cogito Discover, semantic technology platform, is one of the advanced solutions provided by a specialized software company, Expert System® which provides a collection of industrial services span from automation of business processes to social media monitoring and many more. Cogito Discover leverages AI, NLP and Semantic Analytics to bring a consolidated platform that is able to read, analyse and understand multilingual unstructured data, thereby conducting categorization, semantic enrichment, entity and concept extraction and relations interlinking [53]. Figure 5 shows the overall Cogito Discover platform. The depicted model demonstrates its capacity to digest dissimilar data formats and infer enriched data entities and concepts to benefit Business Intelligence applications, customer service support, searching organizational content, to various other domains.

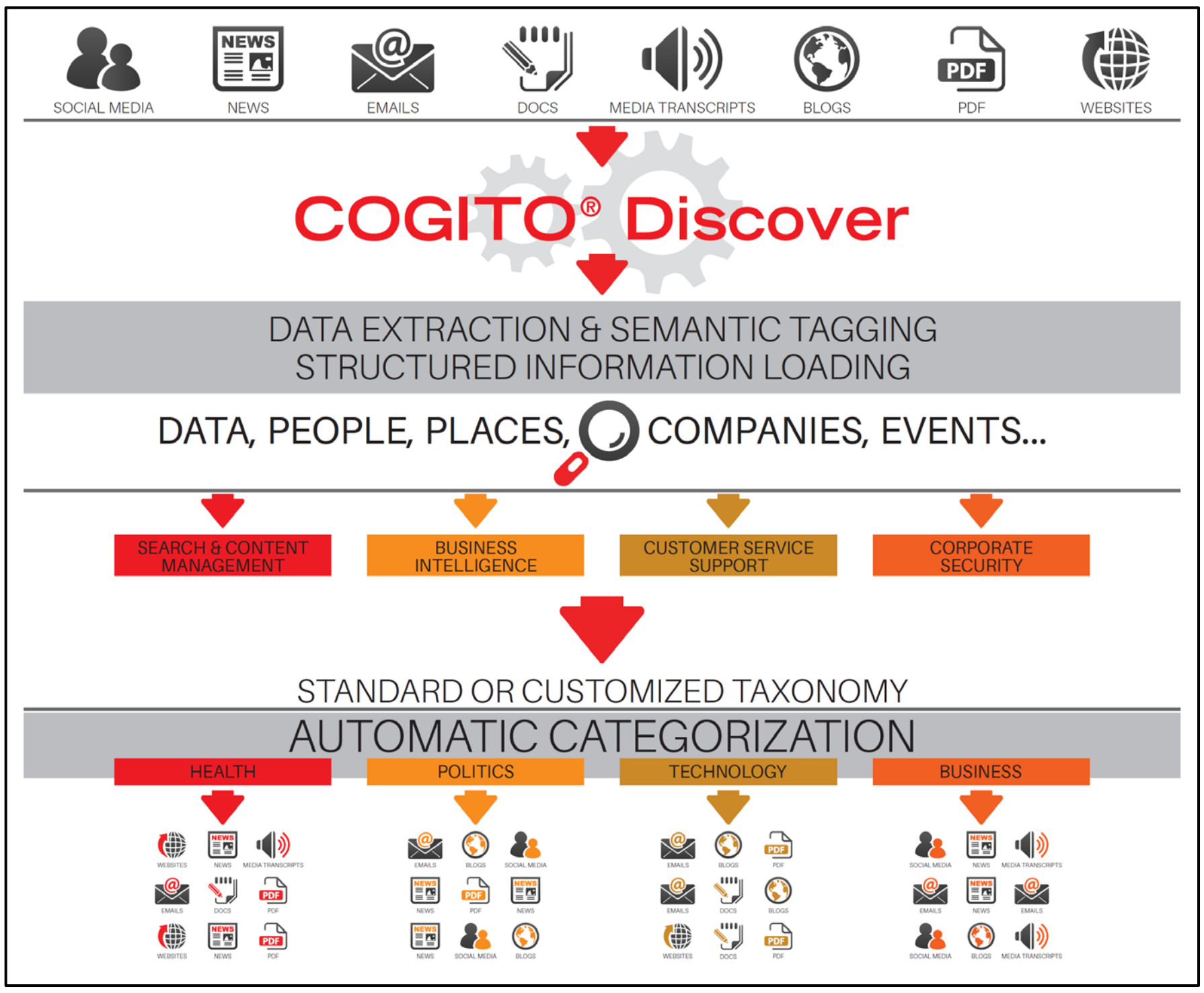

*Figure 5: Cogito Discover Platform [53]*

## 4.5 Knowledge Graphs for SBD

Knowledge Graphs (KGs) have gained considerable attention recently from both academia and industry. Despite the wide usage of KGs in several commercial and industrial applications as well as the incorporation of KG to address several research problems, there has been no consensus on a formal definition of KGs. Also, this emerging venue of research has extended to tackle the continuous propagation of Big data, thereby enabling machines to 'understand' and leverage the domain ontologies to obtain the hoped-for added value. In particular, the proliferation of SBD has prompted the necessity for sophisticated approaches to assist the machine to better understand the context of the multimodal contents. Therefore, the heterogeneity in data sources and format, the discrepancy in vocabulary and the lack of a comprehensive and unified knowledge island are the key challenges for analysts. This section will give an overview of KGs followed by a brief discussion on applications resulted from integrating both social data and KGs.

### 4.5.1 Knowledge Graph – an overview

Knowledge Graph has been used as a core knowledge base in Google's search engine since 2012 to enhance and consolidate retrieved search results with real-life entities collected form widespread range of resources [54]. After coining the term, it has become a buzzword and its application has extended to solve complex problems in numerous domains. The currently proliferated KGs span to embody domain dependent and independent facts. Examples of domain-independent (open-world) KGs include Freebase[8], YAGO[9], Dublin Core (DC[10]), Simple Knowledge Organization System (SKOS[11]), Semantically-Interlinked Online Communities (SIOC[12]), and DBPedia[13] knowledge base, to cite a few. On the other hand, domain-dependent KGs provide an overabundance of benefits to tackle domain-specific problems as well as to acquire the added value from domain corpora [55].

Despite such abundant usage, the term itself has not been given an assented formal definition yet. For example, Ehrlinger et al. [56] attempted to define the term as:

*"A knowledge graph acquires and integrates information into an ontology and applies a reasoner to derive new knowledge."*

Another comprehensive definition that reconciles with the former one was given by Paulheim [57] as follows:

*"A knowledge graph (i) mainly describes real world entities and their interrelations, organized in a graph, (ii) defines possible classes and relations of entities in a schema, (iii) allows for potentially interrelating arbitrary entities with each other and (iv) covers various topical domains."*

More depictions to the term were also elaborated in [58] and [59]. Despite the variance in the definition, there is a notable consensus that KG is constructed on the top of a domain ontology by presenting the domain of knowledge as a set of entities and relations to provide a unified standard representation of data, namely RDF. The Resource Description Framework (RDF) is a widely used underlying model to represent knowledge in terms of triples (*subject , predicate , object*), where subject of the triple indicates the resource which needs to be described, predicates indicates the property of the subject, and object refers to the property value which describes the subject. A typical knowledge graph is represented as a directed graph where nodes indicate the entities (resources) of the class model and edges depicts the relations (properties) between those entities. Figure 6 shows an example of depicted knowledge collected for an Australian politician (Joanne Ryan). As shown in the figure, the knowledge is structured in terms of entities representing real-life classes/concepts and edges which demonstrate the relationships between those entities. The knowledge can then be expressed in terms of triples, namely (subject/head, predicate/relation, tail/object). For example, (Joanne Ryan, wonElection, 2013 Australian Federal Election) is a fact demonstrating the relationship between a person type entity (Joann Ryan) and an event (2013 Australian Federal Election). The relationship is "wonElection" which provides the factual linkage between these two real-life entities. Facts, therefore, presents the heart of the KG, and their embodied entities and predicates can be also extended to describe relationships between instances and/or concepts.

[8] https://developers.google.com/freebase
[9] http://www.foaf-project.org/
[10] https://www.dublincore.org/
[11] http://www.w3.org/2004/02/skos/
[12] http://sioc-project.org/
[13] https://wiki.dbpedia.org/

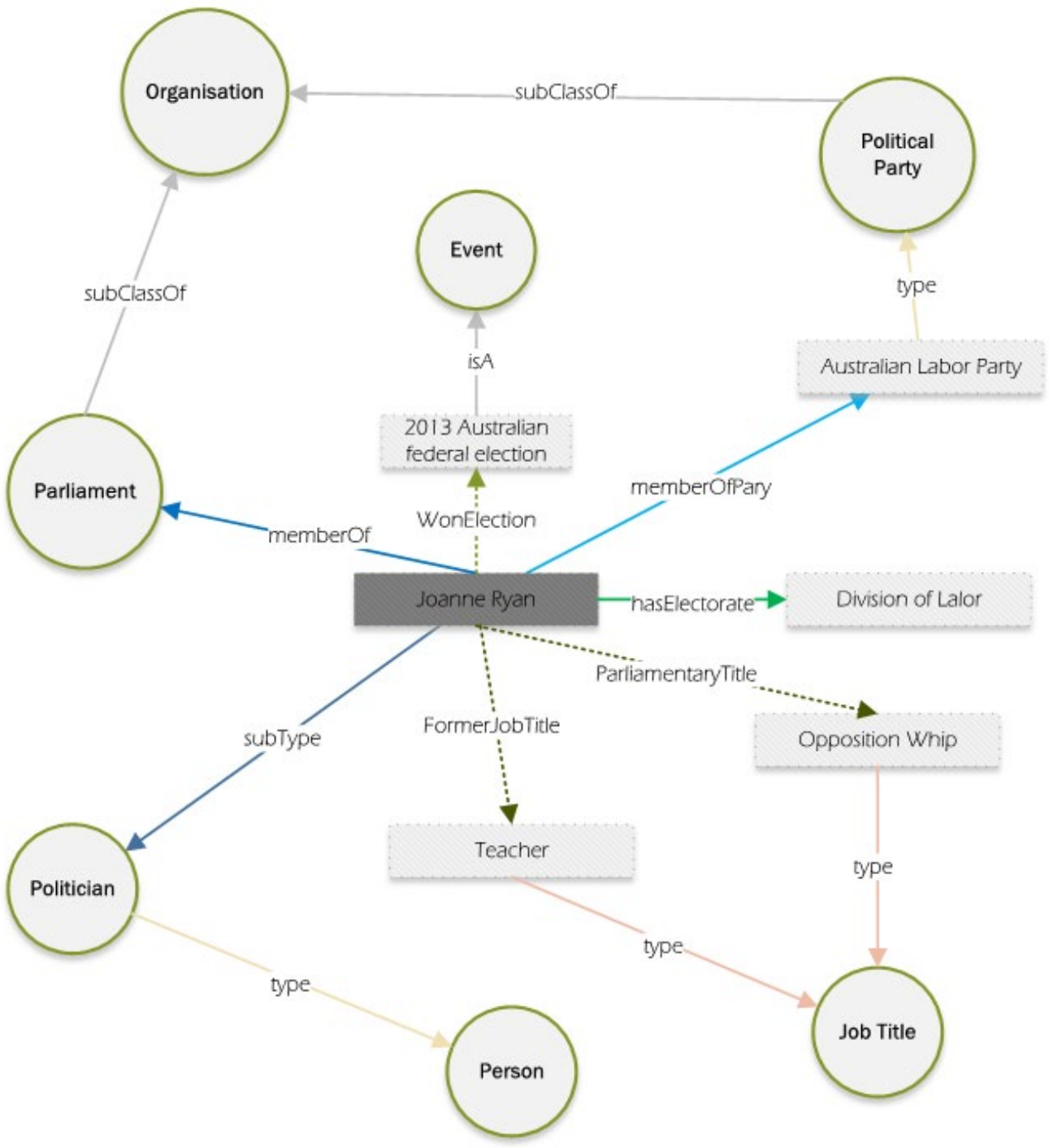

*Figure 6: Knowledge about a politician depicted in terms of entities and relations*

Despite the fact that solving problems pertaining to graphs can be carried out on the conventional graph presentation (i.e. adjacency matrix), mapping the entire graph or its nodes to the vector space has attracted the scientific community due to its scalability to simplify resolving several complex real life graph problems such as KG completion, entity resolution, and link-based clustering, just to cite a few [60-62]. Knowledge Graph Embedding (KGE) is the process of transforming the constituents of a KG (entities and relationships) into a low-dimension semantically-continuous space [63]. Embedding a KG is learned via training a neural architecture over a graph, and comprises commonly three main components, namely; (i) encoding entities into distributed points in the vector space, and encoding relations as vectors, or other forms; (ii) scoring function or model specific function that is used to evaluate the model's efficiency; (iii) optimization procedure, which aims to learn the optimal embedding for the designated KG, thereby the scoring function assigns high scores to positive statements. Further discussion on KG embedding models is out of the scope of this chapter.

### 4.5.2 SBD Applications using KGs:

KGs are commonly constructed from (semi-)structured, such as Wikipedia or unstructured datasets, such as web data using Natural Language Processing (NLP) and linguistic approaches as well as other statistical techniques [57]. KG can be further extended and thereby its embedded knowledge can be augmented to include missing facts of the real world. The process of knowledge acquisition can be categorised into two key dimensions, namely: KG completion, and entity and relation extraction. KG completion aims to expand the current knowledge by accumulating more facts to the current state of the KG, while the latter category aims to infer new knowledge by predicting new relations and entities [64]. Further, link and entity inference in the context of KGs is the process of amplifying the KG with new facts depicted by new entities and/or new relations. SBD provide an extensive resource to amplify domain knowledge. This knowledge can be then used to solve demanding problems and to benefit large scale applications. The following is a selected set of application fields that incorporate SBD to benefit their constructed KGs.

#### 4.5.2.1 Information Retrieval systems

Information Retrieval (IR) allows the storage, management, processing and retrieval of information, documents, websites from dissimilar resources. It is noteworthy that social data has opened a new horizon for IR systems where such user generated contents are a used to consolidate IR embedded engines with wealthy information. This integration has led to what are commonly called Social Information Retrieval models [65]. Therefore, KGs, that have been initially embedded as a core pillar for IR applications, leverage social data contents to offer various IR applications with featured contents. For example, QA systems which are commonly used to provide automated answers to questions incorporating IR techniques and NLP, have been now empowered by KGs and social data to provide better answers to questions. This can be seen in social chatbots and digital applications such as XiaoIce [66]. Research in this direction is still immature, thus further studies should be carried out in this context that can benefit various IR applications such as query expansion, query representation, document expansion personalized indexing, to name a few.

#### 4.5.2.2 Recommender systems

The usage of SBD extends to consolidate recommender systems with knowledge captured from data and metadata collected from users in terms of reviews, posts, comments, likes/dislikes, to name a few. The collected knowledge is used to enrich the knowledge graph with entities and relations that describe real life facts about users and their sentiments toward products and services. The augmented KG can be then utilized to make better and accurate recommendations, find targeted audience and clustering users with similar tastes and interests. Therefore, businesses gain considerable and valuable insights from such augmented KGs that are integrated toward a better brand awareness and visibility, improved customer service and engagement, and discovering new venues for marketing and promoting products and services, etc. [67].

#### 4.5.2.3 Domain Specific applications

**Cybersecurity**: with the ever-increase in quantity and quality of all cybersecurity threats, forces in academia, industry and society are fortified to continuously improving practices, technologies and processes to stop these virtual threats. Several attempts were proposed to make use of KGs as well as machine learning to detect and predict different forms of cyber-attacks and to protect people's cyber assets [67]. Social data offers important means to augment KGs with vast amount of social entities (e.g., individuals, business firm or collective social units) which are networked with dissimilar types of interrelated concepts (e.g., colleagues, friends, family members, people with similar tastes and beliefs) [68]. This additional knowledge can be combined with cybersecurity-specific KGs to enlarge the angle on user behavior, thereby assisting in detecting social patterns related to threats in forms of social engineering, social spam, etc.

**News:** online news services and agencies have been approached by people, organizations, businesses to comprehend what is changing in the world [69]. Furthermore, OSNs provide a new means for people to produce, access and share news in various domains. For example, according to a survey carried out by Pew Research center [70], 62 percent of adults in the United States obtain news from social media. Also, another study conducted by CNN indicated that 75 percent of people receive news by either email or social media posts, and 37 percent of them reshare news through Facebook or Twitter [71]. However, due to the abundance in timely information generated synchronously from these resources as well as the inadequacy to provide users with news that meet their preferences, spreading the news to the targeted audience poses an important challenge [72]. KG and SBD can be combined to address this problem by extracting rich content from users' social data and entities captured from news to model user's preferences. Further, KGs offer a promising area to help the ongoing endeavors to aggregate the unstructured nature of news into a unified structured data store [69].

**Education:** in educational domain, the research on constructing domain KG to benefit stakeholders of this vast domain is far from completeness. Nevertheless, few studies have presented good attempts in this area [73-75]. For example, MOOC (Massive Open Online Courses) providers have started to construct designated KGs to conceptually describing their online resources [76]. In fact, the available educational contents on the Web are increasing exponentially. These contents include both educational-specific resources or generic resources that benefit learners and teachers. In particular, OSNs play a significant role in this sphere by offering students a variety of educational resources as well as offering great opportunities for learners, teachers and even educational institutions to improve their learning methods and practices [77]. Similar to the aforementioned domains, education can positively be influenced from the unique opportunities provided by assimilating social data with the KGs. Thus, further research should be steered in this direction.

## 4.6 A case study on semantic analysis of social politics data

In this section we present a case study on social political data analysis incorporating IBM Watson Natural Language Understanding (NLU) and domain ontology. As indicated previously IBM Watson NLU can identify people, companies, organisations, cities, geographic features, and other types of entities from the textual data content using their built in general classification. Watson NLU supports Linked Data and employs NLP technology to analyse the data and extract the semantically enriched contents. It is a comprehensive ecosystem however it can only categorise the most generic categorization due to the lack of domain-specific knowledge. For a specific domain, Watson NLU will need ontologies to categorise content based on ontology concepts, instances, and relationships. Hence, the ontology-based approach discussed in this section will be of benefit regarding extending the existing Watson NLU.

**Tweet**: "Launched Jennifer Kanis for Melbourne Campaign today. Outcomes instead of ineffective self indulgent commentary. Vote Labor in Melbourne."

**IBM Watson NLU entity extraction and concept mapping results:**

ENTITY: Jennifer Kanis; TYPE of ENTITY: Person
ENTITY: Melbourne Campaign; TYPE of ENTITY: Organization
ENTITY: Melbourne; TYPE of ENTITY: City

**IBM Watson NLU category results:**

/travel/tourist destinations/australia and new zealand

*Figure 7: Output from IBM Watson NLU for entity extraction, concepts mapping, and taxonomy*

As can be seen from Figure 7 , IBM Watson NLU fails to capture the keywords '*vote*' and '*labour*' as entities due to no specific domain knowledge. As a result, the category classifications of travel and society are inadequate. However, if politics ontology is applied as specific domain knowledge, the keywords 'Vote' and 'Labor' are annotated with its type respectively as relation 'voteFor' and concept 'Political Party'. By annotating two more entities of Labor and Vote and specifying particular entity Jennifer Kanis as Politician as shown in Figure 8, the political domain is counted as the domain of this tweet in addition to the travel and society domains. The more data that are annotated, the more entities are extracted in which the domain of tweet is clearer.

**Jennifer Kanis** - CONCEPT: Politician

**Labor** - CONCEPT: Political Party

**Vote** – Relation: voteFor

*Figure 8: Politics Ontology Annotation*

In addition, based on the concepts being referred to, the entities can be inferred to the knowledge captured in the Politics ontology. Figure 9 shows entities ‘Jennifer Kanis’, ‘Labor’, and ‘Vote’ being respectively inferred to concepts ’Politician’ and ‘Political Party’ and relation ‘voteFor’. Figure 9 also shows concepts being related to other concepts forming the domain of knowledge. The knowledge is captured in the politics ontology describes the semantics of concepts. Table 1 shows the modelling notations that appear in Figure 9.

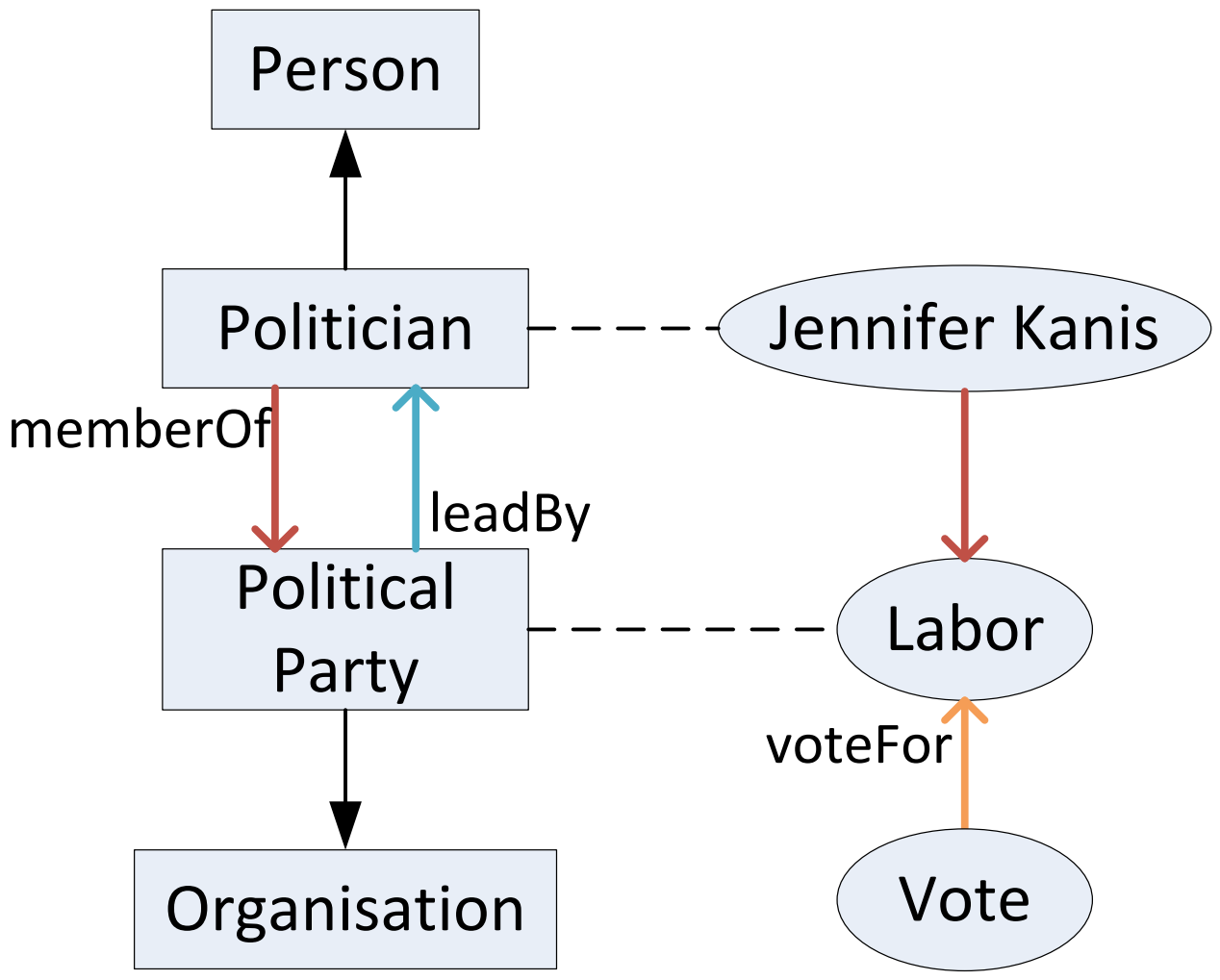

*Figure 9: Knowledge captured in politics ontology*

*Table 1: Ontology modelling notations*

| **Notations** | **Semantics** |
|---|---|
| 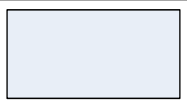 | Concept / Ontology class |
| 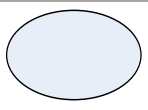 | Instance / Individual |

| Notations | Semantics |
|---|---|
| ⟶ | Association semantical relationship (different colours represent different relationships) |
| ➞ | Generalisation / Taxonomical / Hierarchical relationship |
| - - - | Instance / Individual relationship |

Hence, by integrating the results of the IBM Watson NLU and the politics ontology annotation, the following information can be inferred from the particular tweet:

- Jennifer Kanis is a Politician; Politician is a Person.
- Labor is a Political Party; Political Party is an Organisation.
- Jennifer Kanis is a member of Labor.
- Vote for Labor.
- Melbourne is a city.

Figure 10 indicates the query and subsequent result to retrieve all information of Labour party. As can be seen, it shows entity ‘Labour’ enriched with its type of political party, website, and official name. The entity can also interlink with controlled vocabularies. Here, the entity ‘Labour’ is interlinked with vocabularies from DBpedia, freebase, yago, and semanticweb.

**Query**

```
PREFIX Politics: <http://www.semanticweb.org/ontologies/Politics.owl#>
SELECT *
        WHERE { Politics: labour ?b ?c}
```

**Result**

| b | c |
|---|---|
| http://www.w3.org/1999/02/22-rdf-syntax-ns#type | http://www.semanticweb.org/owl/owlapi/turtle#PoliticalParty |
| http://www.w3.org/2002/07/owl#sameAs | http://dbpedia.org/resource/Australian_Labor_Party |
| http://www.w3.org/2002/07/owl#sameAs | http://rdf.freebase.com/ns/m.0q96 |
| http://www.w3.org/2002/07/owl#sameAs | http://yago-knowledge.org/resource/Australian_Labor_Party |
| http://www.w3.org/2002/07/owl#sameAs | http://www.semanticweb.org/owl/owlapi/turtle#Labor |
| http://www.semanticweb.org/owl/owlapi/turtle#ResolvedName | "Australian Labor Party" |
| http://www.semanticweb.org/owl/owlapi/turtle#Website | "http://www.alp.org.au" |
| http://www.semanticweb.org/owl/owlapi/turtle#value | "labour" |

*Figure 10: Enrichment and interlinking of Labour party*

Figure 11 provides the query that retrieves all information of Politician Daniel Andrews. As can be seen, it shows the enrichment and interlinking of the entity with its name, its type of Politician, and its subclass of Person. The entity is also interlinked with vocabularies from DBpedia, freebase, yago, and semanticweb.

**Query**

```
PREFIX Politics: <http://www.semanticweb.org/ontologies/Politics.owl#>
SELECT *
        WHERE { Politics: DanielAndrews ?p ?o}
```

**Result**

| p | o |
|---|---|
| http://www.w3.org/1999/02/22-rdf-syntax-ns#type | http://www.semanticweb.org/owl/owlapi/turtle#Politician |
| http://www.w3.org/2000/01/rdf-schema#subClassOf | http://www.semanticweb.org/owl/owlapi/turtle#Person |
| http://www.w3.org/2002/07/owl#sameAs | http://dbpedia.org/resource/Daniel_Andrews |
| http://www.w3.org/2002/07/owl#sameAs | http://rdf.freebase.com/ns/m.0bwttx |
| http://www.w3.org/2002/07/owl#sameAs | http://yago-knowledge.org/resource/Daniel_Andrews |
| http://www.w3.org/2002/07/owl#sameAs | http://www.semanticweb.org/owl/owlapi/turtle#DanielAndrews |
| http://www.semanticweb.org/owl/owlapi/turtle#ResolvedName | "Daniel Andrews" |
| http://www.semanticweb.org/owl/owlapi/turtle#value | "danielandrewsmp" |

*Figure 11: Enrichment and interlinking of Politician Daniel Andrews*

Another example is shown in **Error! Reference source not found.**.

**Tweet**: "Thoughts and prayers with Karen Overington's family today. Karen was true Labor, a true friend and will be truly missed by all of us."

**IBM Watson NLU entity extraction and concept mapping results:**

ENTITY: Karen Overington; TYPE of ENTITY: Politician

**IBM Watson NLU category results:**

/society/work/unions
/family and parenting

*Figure 12: Output from IBM Watson NLU for entity extraction, concepts mapping, and taxonomy classification of a tweet*

In the tweet shown in Figure 12 above, IBM Watson NLU captures only the entity 'Karen' Overington as a politician. The entity and keywords of 'true friends', 'prayers', 'thoughts', 'family', and 'labour' are used to classify the tweet under the category of society and family and parenting which is inadequate. Hence, if politics ontology is applied, the keyword 'labour' is annotated as an entity under the concept of the political party. This results in classifying Political domain as an additional domain of tweet.

The politics dataset has been used to evaluate the IBM Watson NLU. IBM Watson NLU classifies the politics dataset into various domains as shown in Figure 13. For two different users, it shows that most tweets are in the travel domain though it is supposed to be in political domain due to the politics dataset. In comparison to results from IBM Watson NLU associated with politics ontology as shown in Figure 14, it classifies the same dataset into the proper domain, i.e. the political domain. This shows significant improvement when associated with specific domain knowledge of politics being captured in politics ontology.

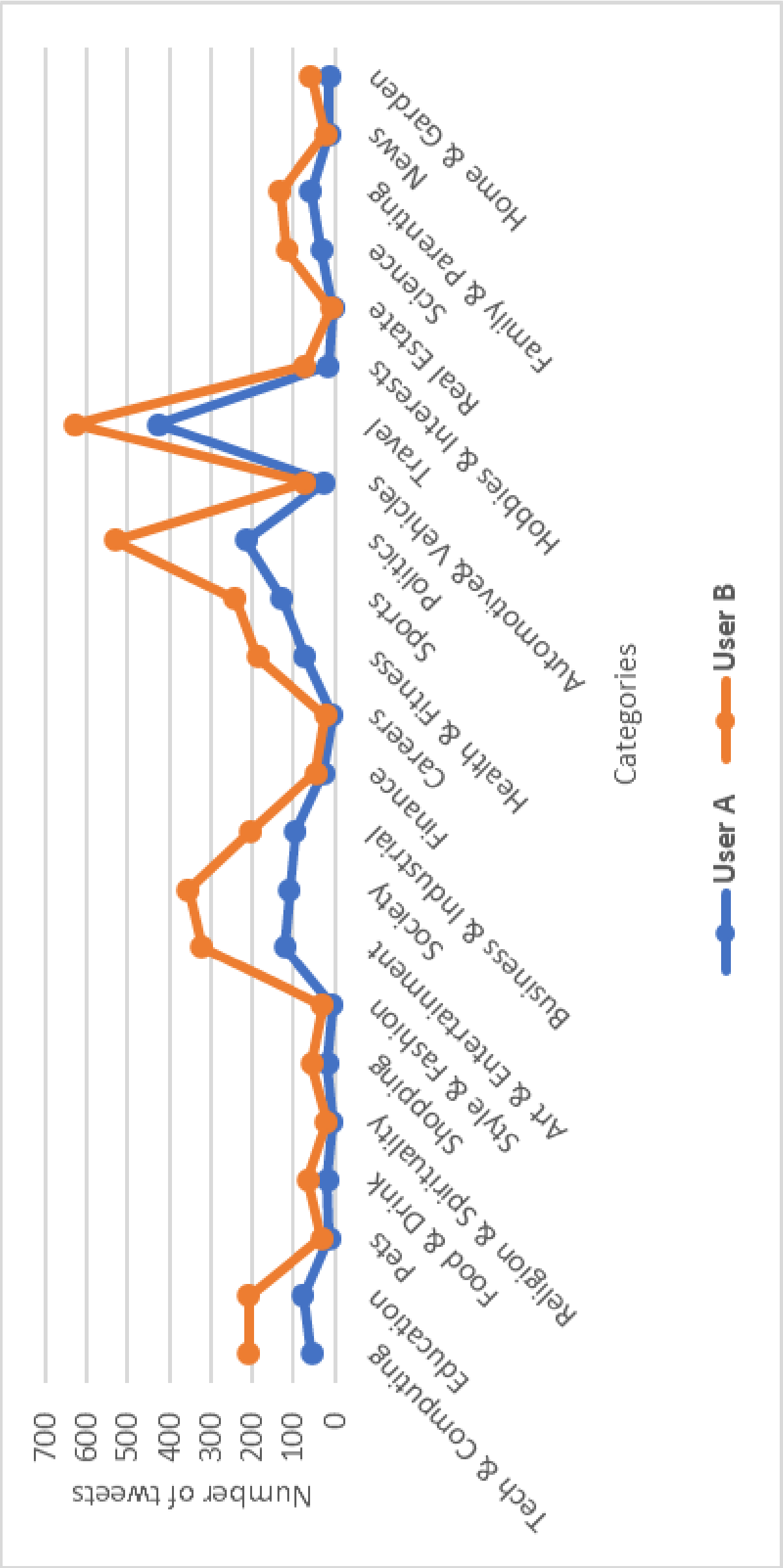

*Figure 13: Results from IBM Watson NLU showing some tweets in various domains from the politics dataset*

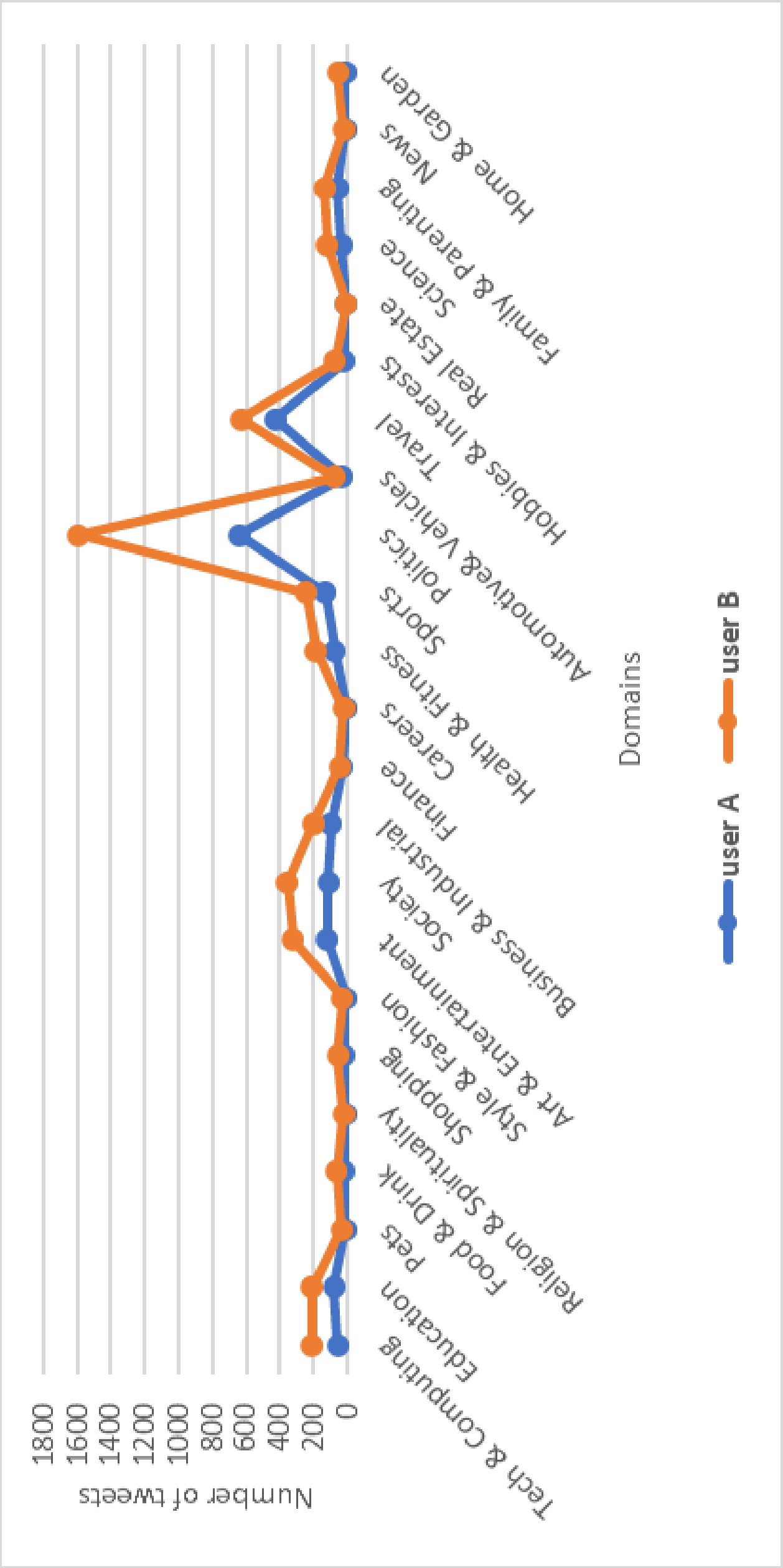

*Figure 14: Results from IBM Watson NLU associated with politics ontology showing a number of tweets in various domains from the politics dataset*

Once the domain can be correctly defined from user's tweets using the ontology-based approach, users' influence in particular domains can be discovered.

## 4.7 Conclusion and Future Works

This chapter discusses semantic analytics and its conjunctions with SBD. In particular, it introduces Ontology, domain knowledge modelling, inference, storage and the related aspects. This chapter also provides a brief on commonly used semantic analysis tools and APIs. Knowledge Graphs are also discussed along with listing some applications that benefit from the integration of Knowledge Graphs and SBD. Finally, a case study incorporating semantic analysis to obtain better understanding to the social content in the context of politics domain is presented and evaluated.

The presented approach for semantic data extraction from SBD has produced optimistic results. It shows a capacity to infer semantically annotated concepts and entities from the textual content leveraged by light-weight ontologies and other semantic repositories. However, the mechanism followed in this approach can be enhanced further:

- IBM Watson NLU has been harnessed and evaluated in the proposed approach as the semantic provider, which has proven to be superior to the other semantic analysis tools. However, in future work, other tools can be utilised and evaluated.

- Twitter user_timeline REST API method has been used to collect public tweets from Twittersphere, which is mainly used to retrieve the historical tweets posted by a certain Twitterer_id. In future work, experiments involving other Twitter APIs will be conducted in order to examine and analyse the textual content related to hashtags and topics mapped, and the latest news and world events and trends.

- Comprehensive ontologies are being continuously updated by applying machine learning technologies, i.e. driving data to obtain the domain knowledge (a reversal of the proposed approach).

- Twitter has been the sole social network on which the experiments and case studies are performed. Hence, a future goal is to analyse other social media networks such as Facebook, LinkedIn, Weblogs etc.

# REFERENCES

1. Sheth, A. and K. Thirunarayan, *Semantics empowered web 3.0: managing enterprise, social, sensor, and cloud-based data and services for advanced applications.* Synthesis Lectures on Data Management, 2012. **4**(6): p. 1-175.
2. Aghaei, S., M.A. Nematbakhsh, and H.K. Farsani, *Evolution of the world wide web: From WEB 1.0 TO WEB 4.0.* International Journal of Web & Semantic Technology, 2012. **3**(1): p. 1-10.
3. Business, F.W. *Past, Present and future outlook of digital technology.* 2020 06/06/2020]; Available from: https://flatworldbusiness.wordpress.com/digital-evolution/.
4. Abu-Salih, B., et al., *Time-aware domain-based social influence prediction.* Journal of Big Data, 2020. **7**(1): p. 10.
5. Abu-Salih, B., et al., *Relational Learning Analysis of Social Politics using Knowledge Graph Embedding.* arXiv preprint arXiv:2006.01626, 2020.
6. Abu-Salih, B., et al. *Social Credibility Incorporating Semantic Analysis and Machine Learning: A Survey of the State-of-the-Art and Future Research Directions*. 2019. Cham: Springer International Publishing.
7. Abu-Salih, B., Alsawalqah, Hamad., Elshqeirat, Basima., Issa, Tomayess., Wongthongtham, Pornpit, *Toward a Knowledge-based Personalised Recommender System for Mobile App Development.* arXiv preprint arXiv:1909.03733, 2019.
8. Hermida, A., et al., *Share, Like, Recommend.* Journalism Studies, 2012. **13**(5-6): p. 815-824.
9. Wongthongtham, P., et al., *State-of-the-Art Ontology Annotation for Personalised Teaching and Learning and Prospects for Smart Learning Recommender Based on Multiple Intelligence and Fuzzy Ontology.* International Journal of Fuzzy Systems, 2018. **20**(4): p. 1357-1372.
10. Wongthongtham, P. and B. Abu-Salih, *Ontology-based approach for identifying the credibility domain in social Big Data.* Journal of Organizational Computing and Electronic Commerce, 2018. **28**(4): p. 354-377.
11. Nabipourshiri, R., B. Abu-Salih, and P. Wongthongtham, *Tree-based Classification to Users' Trustworthiness in OSNs*, in *Proceedings of the 2018 10th International Conference on Computer and Automation Engineering*. 2018, ACM: Brisbane, Australia. p. 190-194.
12. Chan, K.Y., et al., *Affective design using machine learning: a survey and its prospect of conjoining big data.* International Journal of Computer Integrated Manufacturing, 2018. **33**(7): p. 645-669.
13. Chan, K.Y., et al., *Affective design using machine learning: a survey and its prospect of conjoining big data.* International Journal of Computer Integrated Manufacturing, 2018: p. 1-25.
14. Herman, E.S. and N. Chomsky, *Manufacturing consent: The political economy of the mass media*. 2010: Random House.

15. Abu-Salih, B., et al., *CredSaT: Credibility ranking of users in big social data incorporating semantic analysis and temporal factor.* Journal of Information Science, 2018. **45**(2): p. 259-280.
16. Abu-Salih, B., P. Wongthongtham, and K.Y. Chan, *Twitter mining for ontology-based domain discovery incorporating machine learning.* Journal of Knowledge Management, 2018. **22**(5): p. 949-981.
17. Abu-Salih, B., *Trustworthiness in Social Big Data Incorporating Semantic Analysis, Machine Learning and Distributed Data Processing*. 2018, Curtin University.
18. Wongthongtham, P. and B. Abu-Salih. *Ontology and trust based data warehouse in new generation of business intelligence: State-of-the-art, challenges, and opportunities*. in *Industrial Informatics (INDIN), 2015 IEEE 13th International Conference on*. 2015. IEEE.
19. Abu-Salih, B., et al., *An Approach For Time-Aware Domain-Based Analysis of Users' Trustworthiness In Big Social Data.* International Journal of Big Data (IJBD), 2015. **2**(1): p. 16.
20. Abu-Salih, B., et al., *Towards A Methodology for Social Business Intelligence in the era of Big Social Data incorporating Trust and Semantic Analysis*, in *Second International Conference on Advanced Data and Information Engineering (DaEng-2015)*. 2015, Springer: Bali, Indonesia.
21. Corazzon, R., *Ontology. a resource guide for philosophers.* Christian Wolff's "Philosophia prima sive Ontologia"(1729), 2006.
22. Gruber, T.R., *Toward principles for the design of ontologies used for knowledge sharing?* International journal of human-computer studies, 1995. **43**(5): p. 907-928.
23. Stevens, R. *What is an Ontology?* 2001 [cited 2017 3rd March]; Available from: http://www.cs.man.ac.uk/~stevensr/onto/node3.html.
24. Carrasco, R.d.S., et al., *Ontology supported system for searching evidence of wild animals trafficking in social network posts.* Revista Brasileira de Computação Aplicada, 2014. **6**(1): p. 16-31.
25. Iwanaga, I.S.M., et al. *Building an earthquake evacuation ontology from twitter*. in *2011 IEEE International Conference on Granular Computing (GrC)*. 2011.
26. Ghahremanlou, L., W. Sherchan, and J.A. Thom, *Geotagging Twitter Messages in Crisis Management.* The Computer Journal, 2014: p. bxu034.
27. Bontcheva, K. and D. Rout, *Making sense of social media streams through semantics: a survey.* Semantic Web, 2012.
28. Maalej, M., A. Mtibaa, and F. Gargouri, *Ontology-Based Context-Aware Social Networks*, in *New Trends in Databases and Information Systems*, B. Catania, et al., Editors. 2014, Springer International Publishing. p. 287-295.
29. Narayan, S., et al., *Population and Enrichment of Event Ontology using Twitter.* Information Management SPIM 2010, 2010: p. 31.
30. Blei, D.M., A.Y. Ng, and M.I. Jordan, *Latent dirichlet allocation.* Journal of machine Learning research, 2003. **3**(Jan): p. 993-1022.

31. Hofmann, T. *Probabilistic latent semantic indexing*. in *Proceedings of the 22nd annual international ACM SIGIR conference on Research and development in information retrieval*. 1999. ACM.
32. Chen, Y., et al. *Topic modeling for evaluating students' reflective writing: a case study of pre-service teachers' journals*. in *Proceedings of the Sixth International Conference on Learning Analytics & Knowledge*. 2016. ACM.
33. Nichols, L.G., *A topic model approach to measuring interdisciplinarity at the National Science Foundation.* Scientometrics, 2014. **100**(3): p. 741-754.
34. Weng, J., et al. *Twitterrank: finding topic-sensitive influential twitterers*. in *Proceedings of the third ACM international conference on Web search and data mining*. 2010. ACM.
35. Asharaf, S. and Z. Alessandro. *Generating and visualizing topic hierarchies from microblogs: An iterative latent dirichlet allocation approach*. in *Advances in Computing, Communications and Informatics (ICACCI), 2015 International Conference on*. 2015. IEEE.
36. Quercia, D., H. Askham, and J. Crowcroft, *TweetLDA: supervised topic classification and link prediction in Twitter*, in *the 4th Annual ACM Web Science Conference*. 2012, ACM: Evanston, Illinois. p. 247-250.
37. Onan, A., S. Korukoglu, and H. Bulut, *LDA-based Topic Modelling in Text Sentiment Classification: An Empirical Analysis.* Int. J. Comput. Linguistics Appl., 2016. **7**(1): p. 101-119.
38. Michelson, M. and S.A. Macskassy. *Discovering users' topics of interest on twitter: a first look*. in *Proceedings of the fourth workshop on Analytics for noisy unstructured text data*. 2010. ACM.
39. Li, C., et al. *Topic Modeling for Short Texts with Auxiliary Word Embeddings*. in *Proceedings of the 39th International ACM SIGIR conference on Research and Development in Information Retrieval*. 2016. ACM.
40. *BBC Ontologies*. 2015 [cited 2015 19 May]; Available from: http://www.bbc.co.uk/ontologies.
41. Berlanga, R., et al., *Towards a Semantic Data Infrastructure for Social Business Intelligence.* New Trends in Databases and Information Systems, 2014: p. 319-327.
42. Lavalle, S., et al., *Big Data, Analytics and the Path From Insights to Value.* Mit Sloan Management Review, 2011. **52**(2): p. 21-32.
43. Schmidt, B., D. Galar, and L. Wang. *Big Data in Maintenance Decision Support Systems: Aggregation of Disparate Data Types*. in *Euromaintenance 2016 Conference Proceedings*. 2016.
44. Gliozzo, A., et al., *Building Cognitive Applications with IBM Watson Services: Volume 1 Getting Started*. 2017: IBM Redbooks.
45. High, R., *The era of cognitive systems: An inside look at IBM Watson and how it works.* IBM Corporation, Redbooks, 2012: p. 1-16.
46. Redmore, S., *Start from the Question—A Guide to Unstructured Text Analysis.* Business Intelligence Journal, 2015. **20**(4).

47. Salience. *Salience*. 2020 [cited 2020 15/05/2020]; Available from: https://www.lexalytics.com/salience.
48. API, S. *Semantria API*. 2020 [cited 2020 15/05/2020]; Available from: https://www.lexalytics.com/semantria.
49. Lak, P. and O. Turetken. *Star ratings versus sentiment analysis--a comparison of explicit and implicit measures of opinions*. in *2014 47th Hawaii International Conference on System Sciences*. 2014. IEEE.
50. Pampulevski, V., et al., *Sentiment of Media Coverage and Reputation of the Pharmaceutical Industry.* Therapeutic innovation & regulatory science, 2020. **54**(1): p. 220-225.
51. lexalytics. *Top Applications of Sentiment Analysis & Text Analytics*. 2020 [cited 2020 15/05/2020]; Available from: https://www.lexalytics.com/applications.
52. Lexalytics. *Lexalytics Intelligence Platform*. 2020 [cited 2020 15/05/2020]; Available from: https://www.lexalytics.com/platform.
53. System, E. *Cogito Discover*. 2020 [cited 2020 15/05/2020]; Available from: https://expertsystem.com/products/text-analytics-software-cogito-discover/.
54. Singhal, A., *Introducing the knowledge graph: things, not strings.* Official google blog, 2012. **16**.
55. Kejriwal, M., R. Shao, and P. Szekely. *Expert-Guided Entity Extraction using Expressive Rules*. in *Proceedings of the 42nd International ACM SIGIR Conference on Research and Development in Information Retrieval*. 2019.
56. Ehrlinger, L. and W. Wöß, *Towards a Definition of Knowledge Graphs.* SEMANTiCS (Posters, Demos, SuCCESS), 2016. **48**.
57. Paulheim, H., *Knowledge graph refinement: A survey of approaches and evaluation methods.* Semantic web, 2017. **8**(3): p. 489-508.
58. Färber, M., et al., *Linked data quality of dbpedia, freebase, opencyc, wikidata, and yago.* Semantic Web, 2018. **9**(1): p. 77-129.
59. Pujara, J., et al. *Knowledge graph identification*. in *International Semantic Web Conference*. 2013. Springer.
60. Kipf, T.N. and M. Welling, *Semi-supervised classification with graph convolutional networks.* arXiv preprint arXiv:1609.02907, 2016.
61. Wang, X., et al. *Community preserving network embedding*. in *Thirty-first AAAI conference on artificial intelligence*. 2017.
62. Nickel, M., et al., *A review of relational machine learning for knowledge graphs.* Proceedings of the IEEE, 2015. **104**(1): p. 11-33.
63. Wang, Q., et al., *Knowledge graph embedding: A survey of approaches and applications.* IEEE Transactions on Knowledge and Data Engineering, 2017. **29**(12): p. 2724-2743.
64. Ji, S., et al., *A Survey on Knowledge Graphs: Representation, Acquisition and Applications.* arXiv preprint arXiv:2002.00388, 2020.
65. Goh, D. and S. Foo, *Social information retrieval systems: Emerging technologies and applications for searching the web effectively*. 2008: Information Science Reference Hershey, PA.

66. Zhou, L., et al., *The design and implementation of xiaoice, an empathetic social chatbot.* Computational Linguistics, 2020. **46**(1): p. 53-93.
67. Zou, X., *A Survey on Application of Knowledge Graph.*
68. Braun, P., et al., *Knowledge discovery from social graph data.* 2016.
69. Rudnik, C., et al. *Searching News Articles Using an Event Knowledge Graph Leveraged by Wikidata*. in *Companion Proceedings of The 2019 World Wide Web Conference*. 2019.
70. Kohut, A., et al., *In changing news landscape, even television is vulnerable.* Pew Internet & American Life Project, 2012.
71. Gross, D., *Survey: More Americans get news from Internet than newspapers or radio.* CNN, 2010. **1**(3): p. 2010.
72. Liu, D., et al., *News Graph: An Enhanced Knowledge Graph for News Recommendation.*
73. Cui, J. and S. Yu, *Fostering deeper learning in a flipped classroom: Effects of knowledge graphs versus concept maps.* British Journal of Educational Technology, 2019. **50**(5): p. 2308-2328.
74. Shanmukhaa, G.S., S.K. Nandita, and M.V.K. Kiran. *Construction of Knowledge Graphs for video lectures*. in *2020 6th International Conference on Advanced Computing and Communication Systems (ICACCS)*. 2020. IEEE.
75. d'Aquin, M., A. Adamou, and S. Dietze. *Assessing the educational linked data landscape*. in *Proceedings of the 5th Annual ACM Web Science Conference*. 2013.
76. Chen, P., et al., *Knowedu: A system to construct knowledge graph for education.* Ieee Access, 2018. **6**: p. 31553-31563.
77. Tess, P.A., *The role of social media in higher education classes (real and virtual)–A literature review.* Computers in human behavior, 2013. **29**(5): p. A60-A68.